\documentclass[11pt]{article}

\usepackage[latin1]{inputenc}
\usepackage[cyr]{aeguill}
\usepackage[german,english,francais]{babel}

\usepackage{amssymb}

\begin{document}

\begin{flushright}
UCL-IPT-05-02
\end{flushright}

\vspace*{10mm}

\begin{center}
{\Large  Survival before annihilation in $\Psi'$ decays}
\end{center}

\vspace*{5mm}

\begin{center} \baselineskip0.5cm
\textsl{P. Artoisenet,  J.-M. G\'erard and J. Weyers}
\end{center}

\vspace*{3mm}

\begin{center} \baselineskip0.4cm
{\footnotesize   
Institut de Physique Th\'eorique} \\
{\footnotesize   Universit\'e catholique de Louvain}\\
{\footnotesize   B-1348 Louvain-la-Neuve, Belgium}
\end{center}

\vspace*{40mm}

\par\noindent
\textbf{Abstract}

\vspace*{5mm}

{\footnotesize We extend the simple scenario for $\Psi'$ decays suggested a few
years ago. The $c\bar
c$ pair in the $\Psi'$ does not annihilate directly into three gluons but
rather survives before annihilating. An interesting prediction is that a large fraction 
of all $\Psi'$ decays could originate from the $\Psi' \to \eta_{c}
(3\pi)$ channel which we urge experimentalists to identify. Our model
  solves the problem of the apparent hadronic excess in $\Psi'$ decays as well
as the $\rho\pi$ puzzle since, in our view, the two-body decays of the $\Psi'$ are naturally
of electromagnetic origin. Further tests of this picture are proposed, e.g.
$J/\Psi \to b_{1}\eta$.}

\clearpage


\section*{1. Introduction}

The wealth of recent data from BES and CLEO has led to a welcome revival of
interest in charmonium physics. The data now provide an   ideal testing
ground for theoretical expectations on the decay mechanisms at work in the
 $c\bar c$ system.

The conventional picture of a strong three-gluon annihilation of the $\Psi'$
runs into more and more difficulties.  The so-called $\rho \pi$
puzzle and hadronic excess in  $\Psi'$ decays pose indeed challenging problems
\cite{Suz}.

A few years ago, two of us \cite{GerWe} proposed a simple scheme for the decays of
the
$J/\Psi$ and the $\Psi'$. In particular, it was suggested that the $\Psi'$ does
not significantly annihilate into three gluons. In this note, we update and
sharpen the arguments which led to this somewhat unconventional point of view.
In our scenario, all  non-electromagnetic  hadronic decays of the $\Psi'$ have
a simple and general explanation:  survival amplitudes. By this, we mean
transition amplitudes from $\Psi'$ to lower-lying states which still  contain a
$c\bar c$ pair. One original point of this note is the proposal that the
exclusive channel $\Psi' \to
\eta_c + (3\pi)$ could account for a   significant fraction  (possibly
more than $1\%$!) of all $\Psi'$ decays. We do urge our experimental colleagues
to actively search for this forgotten decay mode. 

In Section 2 we briefly review and motivate our point of view on $c \bar c$
annihilation into gluons.  For the $\Psi'$, {\em survival precedes
annihilation}! More precisely, the $c\bar c$ pair survives by spitting out two
or three non-perturbative (i.e. with energy much less than 1 Gev) gluons and
the lower lying pair then annihilates into three or two perturbative (i.e.
with energy  $\gtrsim $ 1 Gev)  gluons, depending on the quantum numbers. These
2+3 or 3+2 annihilation scenarii  get rid of the so-called hadronic
excess problem in $\Psi'$ decays. Many experimental tests of these ideas are  possible. 

To lowest order, survival amplitudes are not expected to hadronize in two-body channels. It
follows that these decays of the $\Psi'$ should result from a direct 
electromagnetic annihilation of the $c\bar c$ pair. Tests and predictions
of this assertion will be discussed in Section 3. In particular, as already
emphasized in our earlier paper, the $\rho \pi$ puzzle is then simply solved.

To conclude this   note we comment very briefly on some other issues in
charmonium physics.

\section*{2.  Strong $c\bar c$ annihilation}

It is well known that in the $J/\Psi$  $(1^{--})$ decays the $c\bar c$ pair
mainly annihilates into three perturbative gluons $(3g)$ or into a photon. The
least massive   channel into which the $3g$  can materialize is $\rho \pi$ which
is indeed the strongest observed hadronic two-body decay \cite{Eid} of the $J/\Psi$.
However, there is also a significant $c\bar c$ survival amplitude namely $J/\Psi
\to \eta_{c}\gamma$. Despite the cost of emitting a photon, this decay has the same
branching ratio as the $\rho \pi$ channel. 

For the $\Psi'$ $(1^{--})$,  the survival radiative decays $\Psi' \to \gamma +
\chi_{c}$ or $\eta_{c}$ $(0^{++}, 1^{++}, 2^{++}$ or $0^{-+})$ are quite
important. Similarly, the dominant strong decay channels have the structure 
\begin{equation}
\Psi' \to  (2NPg) + (3g).
\end{equation}
The physical picture is as simple as   can be: the excited $c\bar c$ pair in
the  $\Psi'$ does not annihilate directly but rather in a two-step process. By
spitting out  two non-perturbative gluons $(2NPg)$,  it first survives in a lower
$c\bar c$ configuration ($1^{--}$ or $1^{+-}$) which then eventually annihilates
into $3g$.  The decays 
$$
  \Psi' \to (2\pi) J/\Psi
\eqno{\mbox{(2a)}}
$$
$$
 \Psi' \to \eta J/\Psi \ \ \ \ 
\eqno{\mbox{(2b)}}
$$
clearly follow the pattern  of Eq. (1). Particularly important from our point of
view is the recently observed \cite{Skw} survival decay 
\setcounter{equation}{2}
\begin{equation}
\Psi' \to \pi^0 h_{c}.
\end{equation}
It is also of the type   Eq. (1) where the $(2NPg)$, the $\eta_{0}$ in this case,
mixes with the
$\pi^0$. The observed rate implies that the effective $\Psi' h_{c} \eta_{0}$
coupling is of the same order as the coupling $\Psi' J/\Psi \eta_{0}$. At
present \cite{Eid},  the survival
radiative decays together with the three on-shell channels (Eqs. (2) and (3))
account for more than 80\% of all $\Psi'$ decays.

The success of the Gell-Mann, Sharp, Wagner off-shell model \cite{Gel} for the
decay  $\omega \to 3\pi$, namely $\omega \to  \pi + ``\rho"$, has led us
\cite{GerWe} to suggest that decay modes, still of the type Eq. (1), 
$$
\begin{array}{llllllll}
&  \hspace*{38mm} & \Psi' \to 2\pi (0^{++})   + ``h_{c}" (1^{+-}) & 
\hspace*{40mm} & \mbox{(4a)}\\
&  \hspace*{38mm} & \Psi' \to \eta (0^{-+})   + ``h_{c}" (1^{+-}) & 
\hspace*{40mm} & \mbox{(4b)}
\end{array}
$$
might also be   the source of sizeable   light hadron decay
modes of the $\Psi'$.   The observed \cite{Eid}
and large $5 \pi$ hadronic decay  of the $\Psi'$ could already correspond to
the pattern of Eq. (4a) where the $``h_{c}"$ is only slightly off-shell.
It
would be nice if, for this decay, experimentalists could identify a two-pion
invariant mass  with the quantum numbers $0^{++}$.

\setcounter{equation}{4}

There is of course  another possibility for a two-step decay pattern 
\begin{equation}
\Psi' \to (3NPg) + (2g) 
\end{equation}
where the lower $c\bar c$ configuration ($0^{-+}$ or $0^{++}$) annihilates
into $2g$.  The only on-shell channel for this type of decays is
$$
\Psi' \to (3\pi)   \eta_{c}
\eqno{\mbox{(6a)}}
$$
to which one may again add the least off-shell amplitude
$$
\Psi' \to 3\pi (1^{--}) +  ``\chi_{c_{0}}" (0^{++}).
\eqno{\mbox{(6b)}}
$$
Eq. (6a) is an original ingredient of this note. It is a genuine survival
amplitude  corresponding to the process where the
$c\bar c$ pair in the $\Psi'$ falls to a lower configuration ($\eta_{c}$) by
radiating three non-perturbative gluons which hadronize in $3\pi$. It could
easily  correspond to $1\%$ or more of all   $\Psi'$ decays. An  effective
calculation (i.e. at the hadronic level) of this decay amplitude 
requires some guesswork about couplings which leaves room for
considerable uncertainty. Details of these calculations will be presented
elsewhere.  It is however quite interesting to point out that the dominant
\cite{Eid} hadronic decay mode  $\Psi' \to 7\pi$ naturally follows from  Eq.
(6a).  We beg experimentalists to search for a $\eta_{c}$ peak in this
multi-pion final state.

In summary, Eqs. (1) and (5) are our explanation of the so-called hadronic
excess in $\Psi'$ decays. If true, there appears to be   no need whatsoever
for an important contribution of direct $\Psi'$ annihilation   into
three gluons. Furthermore, the substitution of one photon for one gluon in Eqs.
(1) and (5) allows \setcounter{equation}{6}
\begin{equation}
\Psi' \to (2NPg) + 2g + \gamma.
\end{equation}
This 2+2+1 pattern corresponds to on-shell radiative decays such as
$$
\Psi' \to (\pi^+ \pi^-) \eta_{c} \gamma
\eqno{\mbox{(8a)}}
$$
$$
\Psi' \to \eta \eta_{c} \gamma \ \ \ \ \ \ \ 
\eqno{\mbox{(8b)}}
$$
which could be larger than the observed $\Psi' \to \eta_{c} \gamma$ mode.

 \section*{3. Electromagnetic $c\bar c$ annihilation}

Whether the $\Psi'$ decays following the 2+3 (Eq. (1)) or 3+2 (Eq. (5))
pattern, it seems intuitively difficult to end up with a light hadronic
two-body channel. This brings us to the suggestion  that these channels are of
electromagnetic origin, namely they follow from the direct hadronization of a
virtual photon. If such is the case, the $e^+e^- \to \gamma^\ast \to$ hadrons
continuum \cite{Wang} should be consistently substracted for {\em all}
two-body branching ratios.

In the   $SU(3)$ limit for hadrons, a well-known consequence of photon
hadronization  is that the ratio of branching ratios into neutral  
and   charged strange states   is expected to be 4 (for $d$-coupling). The recent
data on $\Psi' \to K^\ast \bar K$ agree
very well with this expectation.    Moreover, a  striking difference
between $J/\Psi$ and $\Psi'$ decay modes is observed \cite{Adam}
in the $3\pi$ channel. For the $J/\Psi$, the $\rho$(770) almost saturates the
two-pion invariant mass, while for the $\Psi'$ it is the $\rho$(2150) which 
seems to dominate. Such a strong suppression of the low-lying vector state contributions 
is not surprising in a high-energy electromagnetic process.  These observations
considerably strenghten  the argument that   $\Psi' \to
VP$  is dominantly an electromagnetic process: the so-called $\rho\pi$ puzzle is solved.

\setcounter{equation}{8}

Physically, the $1^{+-}   0^{-+}$ channel $b_{1} \pi$ is even more interesting.
For the moment \cite{Eid}, it is the largest light hadronic two-body decay of
the $\Psi'$
but still of the same order as $\Psi' \to 2(\pi^+\pi^-)$ which is obviously of
electromagnetic nature. If the $b_{1} \pi$ channel comes from the
hadronization of a photon,  then both $\Psi' \to b_{1}$(1235)$\eta$ and $\Psi'
\to h_{1}$(1170)$\pi^0$ should have branching
ratios of the order of $10^{-3}$. These processes are certainly welcome to
saturate the theoretically well-known $\Psi' \to \gamma^\ast \to$ light
hadrons branching ratio  $(\sim 1.6\%)$. Moreover, they lead  to the surprising
prediction that 
\begin{equation}
\mbox{Br} \  (J/\Psi \to h_{1} \pi^0) \thickapprox  \  \mbox{Br} \   (J/\Psi \to
b_{1} \eta) \approx 1\%
\end{equation}
which are of the same order as the measured $J/\Psi \to \rho \pi$ branching
ratio.

 \section*{4. Comments and conclusion}

 The main point of this note has been to reemphasize that there is no
experimental necessity for a direct strong annihilation of the $\Psi'$ into
$3g$. Why is this annihilation process   suppressed? We
can only repeat the argument given earlier \cite{GerWe}: the putative (or
theoretical) $c
\bar c$ states $(n^{2S+1} L_{J})$ are one thing, the physical states are quite
another! With strong annihilation of the $1^{--}$ ground state and its first
``radial excitation",   mixing is expected. The QCD dynamics may
be such that the physical states, presumably mixtures of the putative ones,
are so built up that one of them strongly annihilates into three perturbative
gluons while the other does not. Mixing of the 1 $^{3} S_{1}$ and 2 $^{3}
S_{1}$ states via three perturbative gluons has little effect on the
charmonium mass spectrum, but may be crucial for the decay pattern.
If this explanation is correct, one may wonder
about the decay patterns of the $0^{-+}$ states below the open charm threshold.
For the $\eta'_{c}$, we do expect survival to be significantly more important
than a direct two-gluon annihilation. 

Another comment concerns the 12\% rule: we do not see any reason for this
rule to be valid. Contrary to the electromagnetic annihilation of the $c\bar
c$ into a photon which is a pointlike process, neither the $J/\Psi$ annihilation
into $3g$ nor the 2+3 or 3+2 patterns for the $\Psi'$ are of the same nature
except, possibly, in the $m_{c} \to \infty$ limit, but then sizeable
corrections are to be expected.

To conclude let us repeat that the main points of this short note have been:

1.  to revive and make more precise a very simple picture of {\em all} strong
and radiative  decay modes of the $\Psi'$; 

2. to infer that two-body decays of
$\Psi'$ into light hadrons are of electromagnetic origin.

These simultaneously solve the so-called hadronic excess and $\rho\pi$ puzzle,
respectively. Elegant as this may seem, experimental confirmation is still
required.  The explicit identification of the decays $\Psi' \to \eta_{c} (3\pi),
\eta_{c} (2\pi) \gamma$ and a measurement of the branching ratios for $J/\Psi \to
h_{1} \pi^0, b_{1} \eta$ would be important steps in this direction.

\section*{Acknowledgements}

\vspace*{5mm}
 
This work was supported by the Belgian Federal Office for Scientific,
Technical and Cultural Affairs through the Interuniversity Attraction Pole
P5/27.

\end{document}